\documentclass[11pt]{article}
\usepackage{authblk} 

\usepackage[latin1]{inputenc}
\usepackage[T1]{fontenc}
\usepackage{amsmath}
\usepackage{amsfonts}
\usepackage{amssymb}
\usepackage{enumerate}

\usepackage[normalem]{ulem} 

\usepackage{amsmath,xcolor,ulem}
\usepackage{amsfonts}
\usepackage{amssymb,todonotes}
\usepackage{amsthm}
\usepackage{graphicx}
\usepackage{epsfig}
\usepackage{subcaption}
\usepackage{marginnote}

\usepackage[colorlinks=true, allcolors=blue]{hyperref}

\usepackage[top=2.8cm,bottom=2.8cm,left=2.5cm,right=2.5cm]{geometry}
\usepackage{float}
\usepackage[algoruled]{algorithm2e}
\usepackage[font=footnotesize,width=\linewidth]{caption}

\newcommand{\iu}{\mathrm{i}} % imaginary unit
\DeclareMathOperator{\Op}{Op}
\title{Transparent PT-symmetric nonlinear networks}

\author{M.E.~Akramov$^{1}$, J.R.~Yusupov$^{2}$, M.~Ehrhardt$^{3}$, H.~Susanto$^{4}$ and D.U.~Matrasulov$^{5}$}
\affil{$^1$National University of Uzbekistan, Universitet Str. 4, 100174, Tashkent, Uzbekistan}
\affil{$^2$Kimyo Int. University in Tashkent, 156 Usman Nasyr Str., 100121, Tashkent, Uzbekistan}
\affil{$^3$Bergische Universit\"at Wuppertal, Gau{\ss}strasse 20, D-42119 Wuppertal, Germany}
\affil{$^{4}$Khalifa University, 127788, Abu Dhabi, United Arab Emirates}
\affil{$^5$Turin Polytechnic University in Tashkent, 17
Niyazov Str., 100095, Tashkent, Uzbekistan}

\begin{document}
\maketitle

\begin{tikzpicture}[remember picture,overlay]
	\node[anchor=north east,inner sep=20pt] at (current page.north east)
	{\includegraphics[scale=0.2]{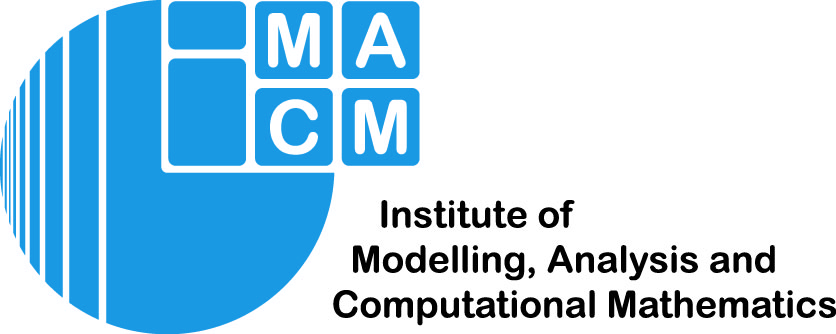}};
\end{tikzpicture}

\begin{abstract}
We consider reflectionless wave propagation in networks modeled in terms of the nonlocal nonlinear Schr\"odinger (NNLS) equation on metric graphs, for which transparent boundary conditions are imposed at the vertices. 
By employing the ``potential approach" previously used for the nonlinear Schr\"odinger equation, we derive transparent boundary conditions for the NNLS equation on metric graphs. 
These conditions eliminate backscattering at graph vertices, which is crucial for minimizing losses in signal, heat, and charge transfer in various applications such as optical fibers, optoelectronic networks, and low-dimensional materials.
\end{abstract}

\begin{minipage}{0.9\linewidth}
 \footnotesize
\textbf{AMS classification:} 65M99, 81-08, 37N20

\medskip

\noindent
\textbf{Keywords:} NNLS equation, metric graphs, transparent boundary conditions, potential approach, nonlinear optics, ferromagnetic structures.
\end{minipage}

%%%%%%%%%%%%%%%%%%%%%%%%%%%%%%%%%%%%%%%%%%%%%%
\section{Introduction}
Nonlocal nonlinear Schr\"odinger (NNLS) equation attracted much attention since its pioneering study by Ablowitz and Muslimani  published in the Ref.~\cite{AM2013}, where they showed integrability of the problem and obtained a soliton solution. 
An interesting feature of the soliton solution of the NNLS equation obtained by Ablowitz and Muslimany is caused by its nonlocality, i.e.\ the solution at a point $x_1$ depends on the solution at point $-x_1$. 
Another important feature is the fact that the NNLS equation is PT-symmetric. 
Later, various aspects of the NNLS equation were studied in a series of papers by Ablowitz and Muslimani \cite{AM2014, AM2016, AM20161, AM2018, AM2018_1, AM2019} and other authors (see, e.g., Refs.~\cite{Stalin, Wen, Yang, Sinha, Hadi2019, Zhenya, Malomed13}). 
Besides nonlocality and PT symmetry, the NNLS equation has practical importance from the  viewpoint  of practical applications in nonlinear optics and some ferromagnetic structures. 
Here we consider the problem of the NNLS equation on metric graphs with a focus on transparent vertex boundary conditions.
The latter means the boundary conditions that ensure the absence of backscattering at the graph vertex. 
To do this, we use the so-called ``potential approach'', which was previously used to impose transparent vertex boundary conditions for the nonlinear Schr\"odinger equation on metric graphs \cite{Jambul1}. 

The motivation for the study of transparent boundary conditions in networks comes from their application in several technologically important problems, such as tunable soliton dynamics in branched optical fibers and optoelectronic networks, and the control of quasiparticle transport in low-dimensional branched functional materials. 
We note that evolution equations on metric graphs have attracted much attention in different context for past two decades \cite{Adami,Zarif,noja,DP2015,Adami16,Vincent1,Adami17,dimarecent,Exciton,BJJEPL,SGN2020}.
In all these cases, it is necessary to reduce losses in signal, heat and charge transfer along the structure by constructing an appropriate network architecture. 

The paper is organized as follows. 
In Section~\ref{sec:2} we briefly introduce soliton solutions and conserving quantities for NNLS equation on a line and recall the main steps of deriving the transparent boundary conditions (TBCs). 
In Section~\ref{sec:3} we derive TBCs for the NNLS equation on metric graphs. 
Section~\ref{sec:4} demonstrates the verification of the obtained results by a numerical experiment. 
Finally, Section~\ref{sec:concl} contains the concluding remarks.

%%%%%%%%%%%%%%%%%%%%%%%%%%%%%%%%%%%%%%
\section{Transparent boundary conditions for the nonlocal nonlinear Schr\"{o}dinger equation on a line}\label{sec:2}

%%%%%%%%%%%%%%%%%%%%%%
\subsection{Soliton solutions of the nonlocal nonlinear Schr\"{o}dinger equation}
Let us consider the NNLS equation on a line % given as
\begin{equation}\label{nnlse}
    \iu\partial_t q(x,t) + \partial^2_x q(x,t) + 2 q(x,t) q^*(-x,t) q(x,t) = 0,
\end{equation}
where $q^*$ denotes the complex conjugate of $q$ and the self-induced potential, 
which can be defined as $V(x,t)=2\,q(x,t)\,q^*(-x,t)$, has the PT-symmetric property, i.e.\ $V(x,t)=V^*(-x,t)$. 
Note that the nonlocality of Eq.~\eqref{nnlse} arises from the term $q^*(-x,t)$ which implies that the solution $q(x,t)$ at coordinate $x$ always requires information from the opposite point $-x$. 
%%%%%%%%%%%%%
For the above NNLS equation, there are many different types of soliton solutions, such as breathing, periodic, rational, and others. 
For example, a single soliton solution found by the inverse scattering method in Ref.~\cite{AM2013} is written as:
\begin{equation}\label{sol01} 
    q(x,t) = -\frac{2(\eta_1+\bar{\eta}_1)\,e^{\iu\bar{\theta}_1}\,e^{4\iu\bar{\eta}^2_1 t}\,e^{-2\bar{\eta}_1x}}{1+e^{\iu(\theta_1+\bar{\theta}_1)} \,e^{-4\iu(\eta^2_1-\bar{\eta}^2_1)t} \,e^{-2(\eta_1+\bar{\eta}_1)x}},
\end{equation}
with $\eta_1$, $\bar{\eta}_1$, $\theta_1$, and $\bar{\theta}_1$ being real constants. 
An important feature of this soliton solution \eqref{sol01} is the fact that it describes a wave that looks like  a ``bird that flaps its wings but does not fly/move".

A traveling soliton solution of Eq.~\eqref{nnlse} derived in Ref.~\cite{Stalin} is written as 
\begin{gather} \label{travelling}
    q(x,t) = \frac{\alpha_1 \,e^{-\Delta/2}\,
    e^{(\bar{\xi}_{1R}-\xi_{1R}) + \iu(\bar{\xi}_{1I}-\xi_{1I})}   }{2\bigl[\cosh(\chi_1)
    \cos(\chi_2) + \iu \sinh(\chi_1)\sin(\chi_2)\bigr]  },
\end{gather}
where 
$\chi_1 = (\xi_{1R}+\bar{\xi}_{1R}+\Delta_R)/2$, 
$\chi_2 = (\xi_{1I}+\bar{\xi}_{1I}+\Delta_I)/2$, 
$\xi_{1R} = -k_{1I} (x+2k_{1R}t)$, 
$\xi_{1I} = k_{1R} x - (k_{1I}^2-k_{1R}^2) t$, 
$\bar{\xi}_{1R} = -\bar{k}_{1I} (x+2\bar{k}_{1R}t)$, 
$\bar{\xi}_{1I} = \bar{k}_{1R} x - (\bar{k}_{1R}^2-\bar{k}_{1I}^2) t$,
\begin{equation*}
\Delta_R = \log\Bigl(\frac{|\alpha_1|^2 |\beta_1|^2}{|k_1+\bar{k}_1|^2}\Bigr),\quad     
\Delta_I = -\frac{\iu}{2} \log\Bigl(\frac{\alpha_1 \beta_1 (k_1^*+\bar{k}_1^*)^2}{\alpha^*_1 \beta^*_1 (k_1+\bar{k}_1)^2} \Bigr), \quad
\Delta = \log{\Bigr(-\frac{\alpha_1 \beta_1}{(k_1+\bar{k}_1)^2}\Bigr)},
\end{equation*}
% \TMatthias{I do not understand the $e^{\Delta}$ here and in (3).}
with $\alpha_1$, $\beta_1$, $k_1$ and $\bar{k}_1$ are arbitrary complex constants, $k_{1R}$, $\bar{k}_{1R}$ and $k_{1I}$, $\bar{k}_{1I}$ are real and imaginary parts of $k_1$, $\bar{k}_1$, respectively.
% \TMashrab{I made a change to the expression of $e^\Delta$.}

The integrability of the problem was proved in \cite{AM2013}, which means that the NNLS equation has infinitely many conservation laws. 
In particular, two important conservation quantities, the norm and the energy, were derived in \cite{AM2013} and are as follows
\begin{equation}\label{energy01}
\begin{split}
    N(t) &= \underset{-\infty}{\overset{+\infty}{\int}}q(x,t)\,q^*(-x,t)\,dx, \\
    E(t) &= \underset{-\infty}{\overset{+\infty}{\int}}\Big[\partial_x q(x,t)\cdot\partial_x q^*(-x,t)+q^2(x,t)\cdot q^{*2}(-x,t)\Big]\,dx.
\end{split}
\end{equation}

The above soliton solutions of Eq.~\eqref{nnlse} are obtained assuming % asymptotic boundary 
decay conditions at infinity, i.e.\ $q(x,t)\to0$ for $x\to\pm\infty$.

%%%%%%%%%%%%%%%%%%%%%%%%%%%%%%%%%%%%%
\subsection{Transparent boundary conditions}
Here, following Ref.~\cite{mashrab}, we briefly recall the problem of transparent boundary conditions (TBCs) for the NNLS equation on a line, which is based on the use of the so-called \textit{potential approach}, 
which was first proposed in \cite{Antoine}. 
The effectiveness of this approach in deriving TBCs for various nonlinear evolution equations has been shown in the Refs.~\cite{Jambul3, TBCSGE, mashrab}. 
Within the framework of this approach, the NNLS equation can be formally reduced to the linear Schr\"{o}dinger equation
\begin{equation}\label{nlse}
   \iu\partial_t q(x,t) + \partial^2_x q(x,t)+ V(x,t) q(x,t)=0, 
\end{equation}
with the potential $V(x,t)=2 q(x,t) q^*(-x,t)$.  
By introducing a new unknown $Q(x,t)$ given by the relation
\begin{equation}\label{Q}
    Q(x,t) = e^{-\iu \mathcal{V}(x,t)}\, q(x,t), 
\end{equation}
with 
\begin{equation}\label{potential}
     \mathcal{V}(x,t) = \int_0^t V(x,s)\,ds,  
\end{equation}
we obtain the Schr\"{o}dinger equation in terms of $Q(x,t)$ as
\begin{equation}\label{L_operator1}
   L(x,t,\partial_x, \partial_t)Q=\iu\partial_t Q + \partial_x^2 Q + A \partial_x Q + B Q = 0, 
\end{equation}
where $A=2\iu\partial_x \mathcal{V}$ and $B=(\iu\partial^2_x \mathcal{V} - (\partial_x \mathcal{V})^2)$.
Using the pseudo-differential operator calculus \cite{taylor} one can linearize Eq.~\eqref{L_operator1} as
\begin{equation}\label{L_operator2}
   L = (\partial_x + \iu\Lambda^-)(\partial_x+\iu\Lambda^+)
     = \partial_x^2 +\iu(\Lambda^+ + \Lambda_-)\partial_x + \iu \Op(\partial_x \lambda^+) - \Lambda^+ \Lambda^-, 
\end{equation}
where $\lambda^+$ is the principal symbol of the operator $\Lambda^+$ and $\Op(p)$ denotes the associated operator of a symbol $p$. 
The Eqs.~\eqref{L_operator1} and \eqref{L_operator2} lead to the system of operators
\begin{equation}
\begin{split}
   & \iu(\Lambda^+ + \Lambda^-) = A, \\
   & \iu \Op(\partial_x \lambda^+) - \Lambda^+ \Lambda^- = \iu\partial_t + B.
\end{split}
\end{equation}
Since the two functions $A$ and $B$ correspond to zero-order operators ($\Op(a)=A$ and $\Op(b)=B$), 
one obtains the symbolic system of equations as
\begin{equation}
\begin{split}
    & \iu(\lambda^+ + \lambda^-) = a, \\
    & \iu\partial_x \lambda^+ -\sum_{\alpha=0}^{+\infty} \frac{(-1)^{\alpha}}{\alpha!} \partial_{\tau}^{\alpha} \lambda^- \partial_t^{\alpha} \lambda^+ = -\tau + b.\label{eq01} 
\end{split}
\end{equation}
 
An asymptotic evolution in the inhomogeneous symbols can be written as
\begin{equation}\label{eq02}
     \lambda^{\pm} \sim \sum_{j=0}^{+\infty} \lambda_{1/2-j/2}^{\pm}. 
\end{equation}
One can determine the $1/2$-order terms in the first relation of the system \eqref{eq01} by substituting the expansion \eqref{eq02} into Eq.~\eqref{eq01}:
\begin{equation}
     \lambda_{1/2}^- = -\lambda_{1/2}^+, \quad \lambda_{1/2}^+ = \pm \sqrt{-\tau}.
\end{equation}
Here, the choice $\lambda_{1/2}^+ = \pm \sqrt{-\tau}$ corresponds to the Dirichlet-to-Neumann (DtN) operator. 
The system of equations for the zeroth-order terms can be written as
\begin{equation}
\begin{split}
  & \lambda_0^- = -\lambda_0^+ -\iu a, \\
  & \iu\partial_x \lambda_{1/2}^+ - (\lambda_0^- \lambda_{1/2}^+ + \lambda_0^+ \lambda_{1/2}^-)=0.\label{eq03} 
\end{split}
\end{equation}
Then, from Eq.~\eqref{eq03} we obtain
\begin{equation}
\begin{split}
  \lambda_0^+ &= -\iu\frac{a}{2} = \frac{1}{2} \partial_x \mathcal{V}, \\
  \lambda_0^- &= -\lambda_0^+ -\iu a = \frac{1}{2} \partial_x \mathcal{V}.
\end{split}
\end{equation}
Since $\partial_t^{\alpha} \lambda_{-1/2}^{\pm} = \partial_{\tau}^{\alpha} \lambda_0^{\pm}=0$, $\alpha\in N$, for the terms of order $-1/2$ we get
\begin{equation}
\begin{split}
    & \iu(\lambda_{-1/2}^+ + \lambda_{-1/2}^-) =  0, \\
    & \iu\partial_x \lambda_0^+ - (\lambda_{-1/2}^- \lambda_{1/2}^+ + \lambda_0^+ \lambda_0^- + \lambda_{-1/2}^+ \lambda_{1/2}^-) = b.\label{eq04}
\end{split}
\end{equation}
From Eq.~\eqref{eq04} we obtain
\begin{equation}
   \lambda_{-1/2}^{\pm}=0.
\end{equation}
In the same way one can obtain the terms of the next order as
\begin{equation}
    \lambda_{-1}^- = -\lambda_{-1}^+ ,
    \quad \lambda_{-1}^+ = \iu\frac{\partial_x V}{4\tau}.
\end{equation}
As a result, TBCs were derived up to the second-order approximation:
\begin{subequations}\label{second_approx}
\begin{align} 
    & \partial_x q\big|_{x=-L} - e^{-\iu\frac{\pi}{4}} e^{\iu\mathcal{V}} \partial^{1/2}_t (e^{-\iu\mathcal{V}}q) \big|_{x=-L} 
    - \iu\frac{\partial_x V}{4} e^{\iu\mathcal{V}} I_t(e^{-\iu\mathcal{V}}q) \big|_{x=-L} = 0, \label{second_approx1}\\
    & \partial_x q\big|_{x=L} + e^{-\iu\frac{\pi}{4}} e^{\iu\mathcal{V}} \partial^{1/2}_t (e^{-\iu\mathcal{V}}q) \big|_{x=L}  + \iu\frac{\partial_x V}{4} e^{\iu\mathcal{V}} I_t(e^{-\iu\mathcal{V}}q) \big|_{x=L}=0.\label{second_approx2}
\end{align}
\end{subequations}
where the operator $\partial^{1/2}_t$, which denotes the half-order fractional time derivative operator, is defined as
\begin{equation}\label{fractder}
   (\partial_t^{1/2} f) (t) = \frac{1}{\sqrt{\pi}} \partial_t \int_0^t \frac{f(s)}{\sqrt{t-s}}\,ds,
\end{equation}
and the operator $I_t(f)$ is
\begin{equation}\label{intf}
   (I_t f)(t)= \int_0^t f(s) \,ds.
\end{equation}

%%%%%%%%%%%%%%%%%%%%%%%%%%%%%%%%%%%%%%%%%%%%%%
\section{Transparent boundary conditions for the nonlocal nonlinear Schr\"{o}\-dinger equation on metric graphs}\label{sec:3}

%%%%%%%%%%%%%%%%%%%%%%%%%%%%%%%%%
\subsection{Nonlocal nonlinear Schr\"odinger equation on a star graph}\label{sec:3A}

\begin{figure}[t!]
\includegraphics[width=95mm]{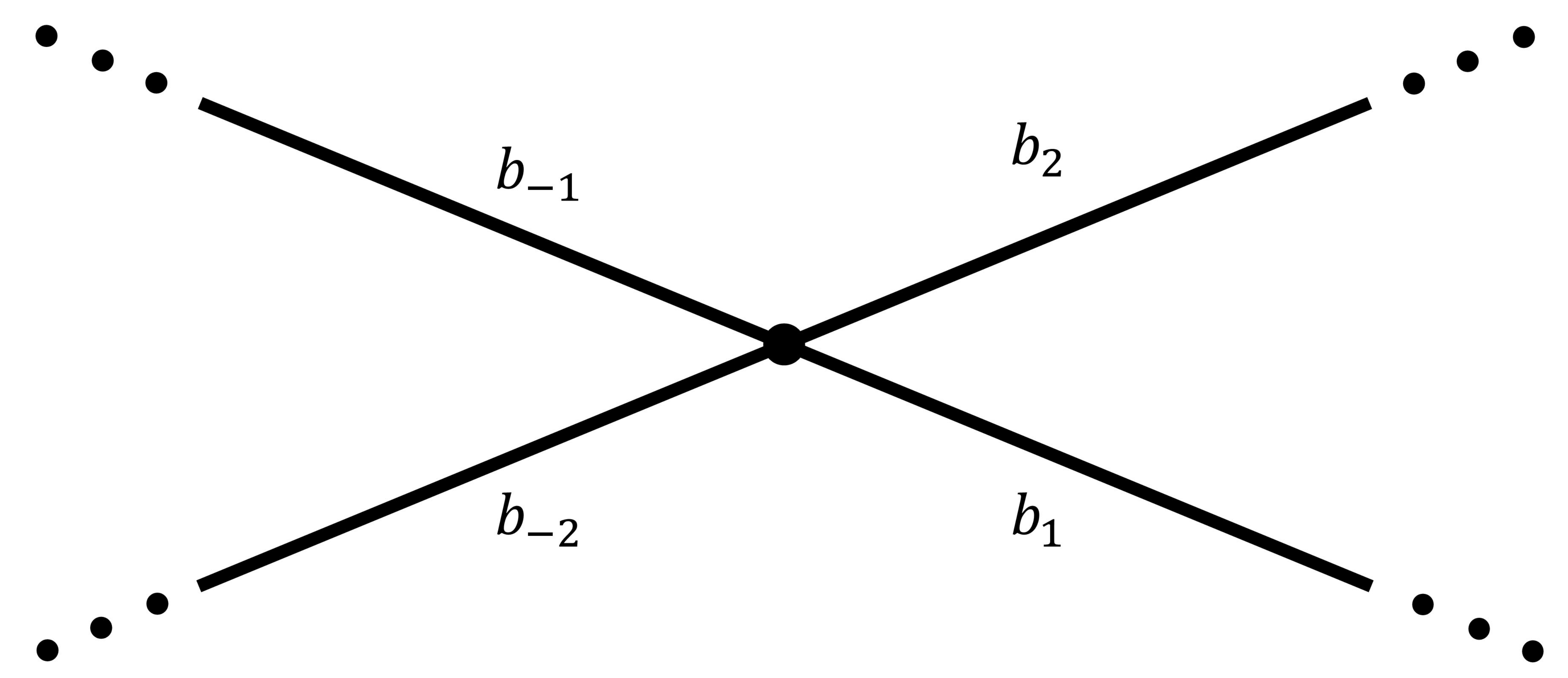}
\caption{The simplest symmetric star graph with four bonds.} \label{fig:star_graph}
\end{figure}

One of the restrictions on the class of initial conditions for NNLSE is that they must be even (in $x$), which makes the initial data symmetric with respect to the $y$-axis.
Taking this property into account, we consider the simplest possible star graph with an even number of bonds (see, Fig.~\ref{fig:star_graph}). 
The nonlocal nonlinear Schr\"odinger equation is written on each bond of the star graph with bonds $b_{\pm j}$ to which a coordinate $x_{\pm j}$ is assigned.
We choose the origin of the coordinates at the vertex so that the bond $b_{-j}$ takes values $x_{-j}\in (-\infty,0)$ and for $b_j$ we fix $x_j\in[0,+\infty)$:
\begin{gather}\label{nnlse1}
 \iu\partial_t q_{\pm j}(x,t)+\partial^2_x q_{\pm j}(x,t) + \sqrt{\beta_{j}\beta_{-j}}\,
 q_{\pm j}^2(x,t)q_{\mp j}^* (-x,t) = 0,
\end{gather}
where $q_{\pm j}(x,t)$ are defined in $x\in b_{\pm j}$, and $j=1,2$.

The Eq.~\eqref{nnlse1} is a system of NNLS equations where components of $q_{\pm j}$ are mixed in the nonlinear term due to the presence of the factor $\sqrt{\beta_{j}\beta_{-j}}$. 
To solve this equation, it is necessary to impose boundary conditions at the branching point (vertex) of the graph. 
Here we choose the boundary conditions derived in \cite{Mashrab2022}, which ensure that the considered system is integrable, and they are derived by showing that there exists an infinite number of conservation laws. 
Within this approach for the above NNLSE, the norm is determined as, cf.\ \cite{AM2013}
\begin{equation}\label{norm1}
  N(t) = \sum_{j=1}^2 \bigl[ N_j(t) + N_{-j}(t) \bigr], \quad 
          N_{\pm j}(t)=\int_{b_{\pm j}}q_{\pm j}(x,t)q_{\mp j}^*(-x,t)\,dx.
\end{equation}
Another conserving quantity, i.e., the energy, is given by
\begin{multline}\label{energy1}
   E(t) = \sum_{j=1}^2 \bigl[ E_j(t) + E_{-j}(t) \bigr], \quad 
          E_{\pm j}(t)=\int_{b_{\pm j}} 
\Bigl(\partial_x q_{\pm j}(x,t)\cdot \partial_x q_{\mp j}^*(-x,t)+\frac{\sqrt{\beta_{j} \beta_{-j} }}{2} q_{\pm j}^2(x,t)\cdot q_{\mp j}^{*2}(-x,t)\Bigr)\,dx.  
\end{multline}
By requiring the conservation of these quantities, the time derivatives of the norm and the energy lead to the following vertex boundary conditions \cite{Mashrab2022}:
\begin{equation}
\begin{split}
  & \gamma_{1} q_{1}(x,t)\big|_{x=0} = \gamma_{-1} q_{-1}(x,t)\big|_{x=0} 
   =\gamma_{2} q_{2}(x,t)\big|_{x=0} = \gamma_{-2} q_{-2}(x,t)\big|_{x=0}, \\
  & \frac{1}{\gamma_{1}}\partial_x q_{1}(x,t)\big|_{x=0} 
  + \frac{1}{\gamma_{2}}\partial_x q_{2}(x,t)\big|_{x=0}=
   \frac{1}{\gamma_{-1}}\partial_x q_{-1}(x,t)\big|_{x=0}+
   \frac{1}{\gamma_{-2}}\partial_x q_{-2}(x,t)\big|_{x=0},    \label{bc1}
\end{split}
\end{equation}
where the parameters $\gamma_{\pm j}$ are non-zero positive real numbers.

Then the solution of the problem given by Eqs.~\eqref{nnlse1} and \eqref{bc1} can be expressed in terms of the solution of Eq.~\eqref{nnlse} as 
\begin{equation}
    q_{\pm j}(x,t) = \sqrt{\frac{2}{\beta_{\pm j}}}\,q(x,t),
\label{solit}
\end{equation}
and it satisfies the boundary conditions \eqref{bc1}, provided that the following conditions hold:
\begin{gather}
  \frac{\gamma_{\pm j}}{\gamma_{-1}}=\sqrt{\frac{\beta_{\pm j}}{\beta_{-1}}},\quad\quad
  \frac{1}{\beta_{1}}+\frac{1}{\beta_{2}}=
  \frac{1}{\beta_{-1}}+\frac{1}{\beta_{-2}}.\label{constrain1}
\end{gather}

A traveling soliton solution of Eq.~\eqref{travelling} given on a graph can be written as
\begin{gather} \label{travelling_graph}
    q_{\pm j}(x,t) = \sqrt{\frac{2}{\beta_{\pm j}}} 
    \frac{\alpha_1 \,e^{-\Delta/2}\,
    e^{(\bar{\xi}_{1R}-\xi_{1R}) + \iu(\bar{\xi}_{1I}-\xi_{1I})}   }
    {2\bigl[\cosh(\chi_1)\cos(\chi_2) + \iu \sinh(\chi_1)\sin(\chi_2)\bigr]  }.
\end{gather}

The sum rule \eqref{constrain1} can be considered as a condition (constraint) that ensures the integrability of NNLS equation on a metric star graph given by Eqs.~\eqref{nnlse1} and \eqref{bc1}. 
In other words, if the sum rule \eqref{constrain1} is fulfilled,
there exist an analytical solution which can be expressed as Eq.~\eqref{travelling_graph}.

%%%%%%%%%%%%%%%%%%%%%%%%%%%%%%%%%%%%%%%%%%%%%%%%%%%%%%%%%
\subsection{Derivation of transparent vertex boundary conditions}\label{sec:3B}

In this subsection, we derive the TBCs for the nonlocal nonlinear Schr\"odinger equation on graphs by applying the potential approach used in the derivation of TBCs on a line. 
Subsequently, the NNLS equation can be formally written as a linear PDE 
\begin{equation}\label{nnlse_potential}
     \iu\partial_t q_{\pm j}(x,t)+\partial^2_x q_{\pm j}(x,t) + V_{\pm j}(x,t) q_{\pm j}(x,t) =0,
\end{equation}
where $V_{\pm j}(x,t)=\sqrt{\beta_{j}\beta_{-j}}\,q_{\pm j}(x,t) q^*_{\mp j}(-x,t)$.

Now we split the whole domain (graph) into two subdomains, which we call ``interior'' (bonds $b_{\pm 1}$) and ``exterior" (bonds $b_{\pm 2}$). 
We use these terminologies to be consistent with those that were used for the problem considered on a line. 
Moreover, the terminologies are borrowed from the original works \cite{Ehrhardt1999, Ehrhardt2001, Arnold2003, Antoine2008}, in which the basic idea of constructing TBCs was proposed.
Accordingly, we consider in the sequel interior and exterior problems. 
The interior problem for $b_{\pm 1}$ can be written as
\begin{equation}    
\begin{split}
  &\iu\partial_t q_{\pm 1} +\partial_x^2 q_{\pm 1} + V_{\pm 1}(x,t) q_{\pm 1} =0,\label{intprob}\\
  &q_{\pm 1}\big|_{t=0}=Q^I(x), \\
  &\partial_x q_{\pm 1}\big|_{x=0}=\pm (T_0 q_{\pm 1})\big|_{x=0}, 
\end{split}
\end{equation}
where $Q^I(x)$ is an initial condition and $T_0$ is yet an unknown operator that determines the TBCs.

%%%%%%%%%%%%%%%%%%
The exterior problems for $b_{\pm 2}$ reads
\begin{equation}
\begin{split}
   &\iu\partial_t q_{\pm 2} + \partial_x^2 q_{\pm 2} + V_{\pm 2}(x,t) q_{\pm 2}=0, \label{extprob} \\
   &q_{\pm 2}\big|_{t=0} = 0, \\
   &q_{\pm 2}\big|_{x=0} = \psi_{\pm 2}(t), \quad \psi_{\pm 2}(0)=0, \\
   &(T_0\psi_{\pm 2})\big|_{x=0}=\mp\partial_x q_{\pm 2}\big|_{x=0}. 
\end{split}
\end{equation}

%%%%%%%%%%%%%%%%
We introduce a new function
\begin{equation}\label{eq:vf1}
    \mu_{\pm j} (x,t)=e^{-\iu\nu_{\pm j} (x,t)} q_{\pm j}(x,t),
\end{equation}
where
\begin{equation}
    \nu_{\pm j}(x,t) =\int_0^t V_{\pm j}(x,\tau)\,d\tau 
    =\sqrt{\beta_{j}\beta_{-j}}\int_0^t q_{\pm j}(x,\tau) q^*_{\mp j}(-x,\tau)\,d\tau.\label{eq:nu2}
\end{equation}

Then, the TBCs of the second order approximation \eqref{second_approx} for $q_{\pm 2}$ at $x=0$ can be written as 
\begin{equation}\label{tbc_qpm2}
    \partial_x q_{\pm 2}\big|_{x=0} = \pm e^{-\iu\frac{\pi}{4}}e^{\iu\nu_{\pm 2}}\cdot\partial_t^{1/2}(e^{-\iu\nu_{\pm 2}} q_{\pm 2})\big|_{x=0}
     \pm \iu\frac{1}{4}\partial_x V_{\pm 2}e^{\iu\nu_{\pm 2}}I_t(e^{-\iu\nu_{\pm 2}} q_{\pm 2})\big|_{x=0},
\end{equation}
where the fractional 1/2-derivative and $I_t$ are given by \eqref{fractder} and \eqref{intf}, correspondingly.

Thus, we find the $T_0$ operator for $q_{\pm j}$ for $x=0$ as
\begin{equation}\label{tbc_qpm3}
    (T_0q_{\pm j})\big|_{x=0} = - e^{-\iu\frac{\pi}{4}}e^{\iu\nu_{\pm 2}}\cdot\partial_t^{1/2}(e^{-\iu\nu_{\pm j}} q_{\pm j})\big|_{x=0}
     - \iu\frac{1}{4}\partial_x V_{\pm j}e^{\iu\nu_{\pm j}}I_t(e^{-\iu\nu_{\pm j}} q_{\pm j})\big|_{x=0}.
\end{equation}
To find the TBC for $q_{\pm 1}$ at $x=0$, we apply operator $T_0$ to $q_{\pm 1}$ as
\begin{equation}
    \partial_x q_{\pm 1}|_{x=0} = \mp e^{-\iu\frac{\pi}{4}}e^{\iu\nu_{\pm 2}}\cdot\partial_t^{1/2}(e^{-\iu\nu_{\pm 2}} q_{\pm 2})\big|_{x=0}
     \mp \iu\frac{1}{4}\partial_x V_{\pm 2}e^{\iu\nu_{\pm 2}}I_t(e^{-\iu\nu_{\pm 2}} q_{\pm 2})\big|_{x=0}. \label{tbc_qpm1}
\end{equation}

From the continuity of the solution in Eq.~\eqref{bc1} we have
\begin{equation} \label{eq2}
\begin{split}
    &\nu_{-1}(0,t)=\nu_{-2}(0,t)=\nu_{1}(0,t)=\nu_{2}(0,t),\\
    & V_{-1}(0,t)=V_{-2}(0,t)=V_1(0,t)=V_2(0,t),\\
    & \sqrt{\beta_{-1}} (T_0 q_{-1})\big|_{x=0} = 
    \sqrt{\beta_{1}} (T_0 q_{1})\big|_{x=0} = 
    \sqrt{\beta_{-2}} (T_0 q_{-2})\big|_{x=0} =
    \sqrt{\beta_{2}} (T_0 q_{2})\big|_{x=0}.
\end{split}
\end{equation}
% which leads to
% \begin{align}
%     & \partial_x q_{1}|_{x=0} = 
%     -\sqrt{\frac{\beta_{-1}}{\beta_1}} \left[ e^{-\iu\frac{\pi}{4}}e^{\iu\nu_{-1}}\cdot\partial_t^{1/2}(e^{-\iu\nu_{-1}} q_{-1})|_{x=0}
%      +\iu\frac{1}{4}\partial_x V_{-1}e^{\iu\nu_{-1}}I_t(e^{-\iu\nu_{-1}} q_{-1})|_{x=0}=0 \right], \\
%      & \partial_x q_{\pm 2}|_{x=0} = 
%     \pm \sqrt{\frac{\beta_{-1}}{\beta_{\pm 2}}} \left[ e^{-\iu\frac{\pi}{4}}e^{\iu\nu_{-1}}\cdot\partial_t^{1/2}(e^{-\iu\nu_{-1}} q_{-1})|_{x=0}
%      +\iu\frac{1}{4}\partial_x V_{-1}e^{\iu\nu_{-1}}I_t(e^{-\iu\nu_{-1}} q_{-1})|_{x=0}=0 \right],
% \end{align}

% \begin{figure}[t!]
% \includegraphics[width=150mm]{fulfilled.pdf}
% \caption{Soliton dynamics at different time moments, i.e., $t=0$ (a), $t=0.8$ (b), $t=1.1$ (c) and $t=2$ 
% (d) when the sum rule is fulfilled by choosing the following values of the nonlinearity coefficients: $\beta_{\pm j}=2$.} \label{fig:sol_dyn}
% \end{figure}

% ------- Sum rule 1 is fulfilled in Eq 26, i.e., norm conserved case
\begin{figure}[t!]

\begin{subfigure}{.475\linewidth}
  \includegraphics[width=\linewidth]{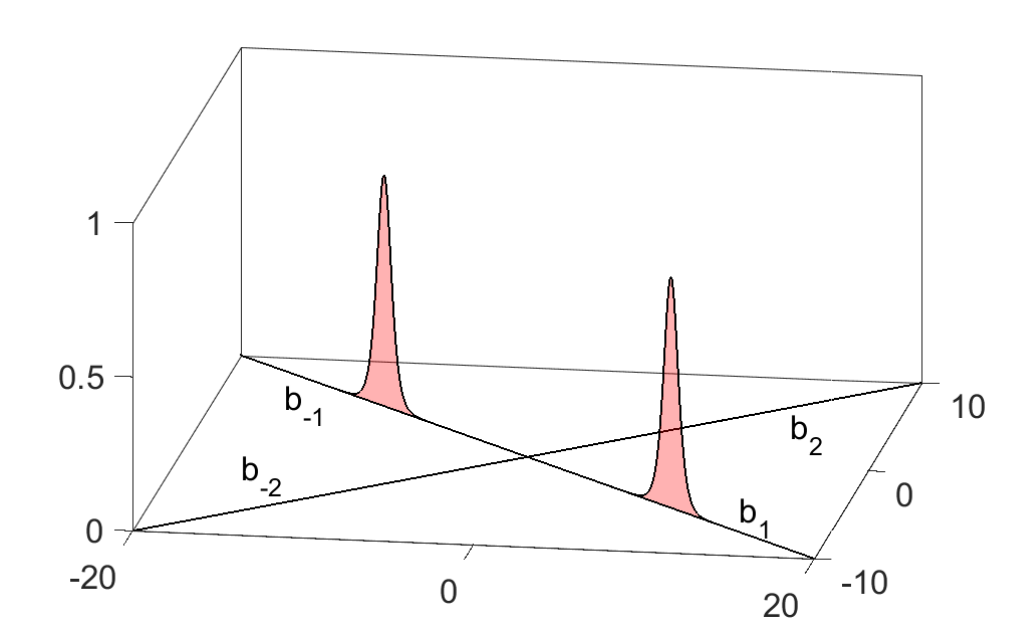}
  \caption{}
  % \label{MLEDdet}
\end{subfigure}\hfill % <-- "\hfill"
\begin{subfigure}{.475\linewidth}
  \includegraphics[width=\linewidth]{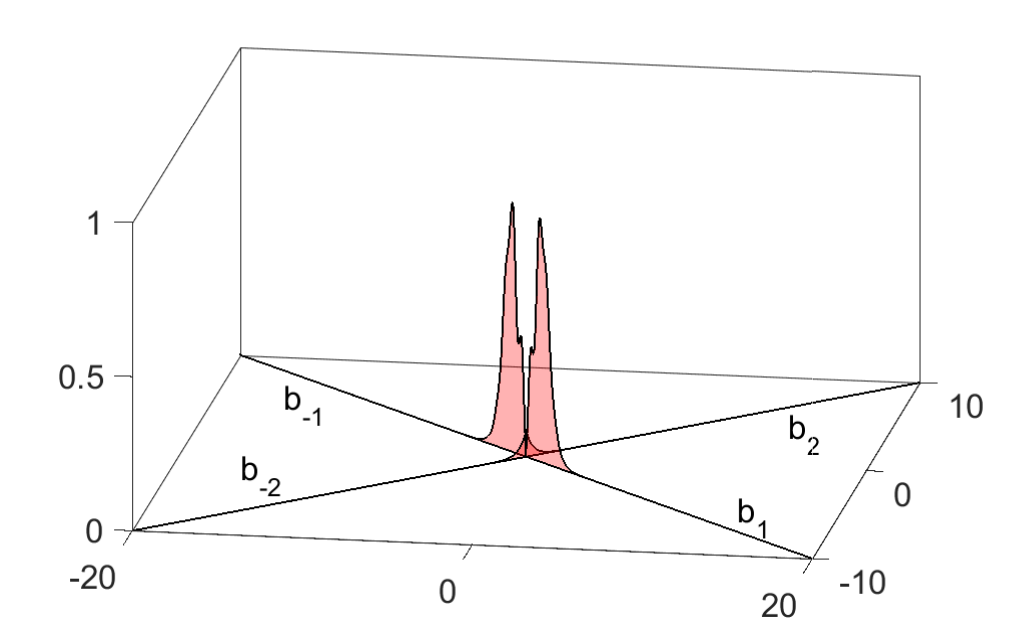}
  \caption{}
  % \label{energydetPSK}
\end{subfigure}

\medskip % create some *vertical* separation between the graphs
\begin{subfigure}{.475\linewidth}
  \includegraphics[width=\linewidth]{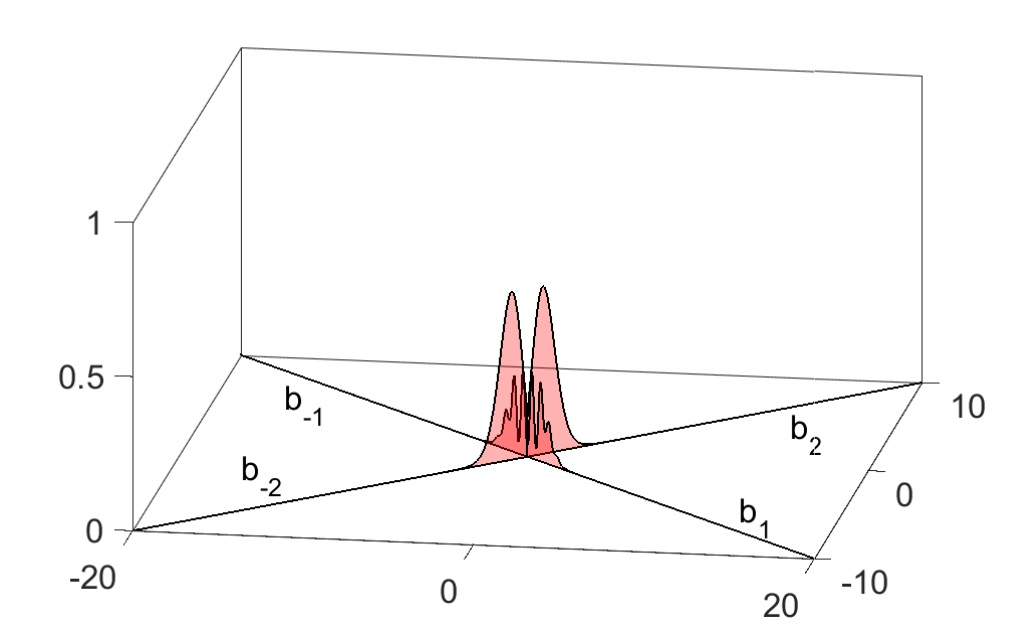}
  \caption{}
  % \label{velcomp}
\end{subfigure}\hfill % <-- "\hfill"
\begin{subfigure}{.475\linewidth}
  \includegraphics[width=\linewidth]{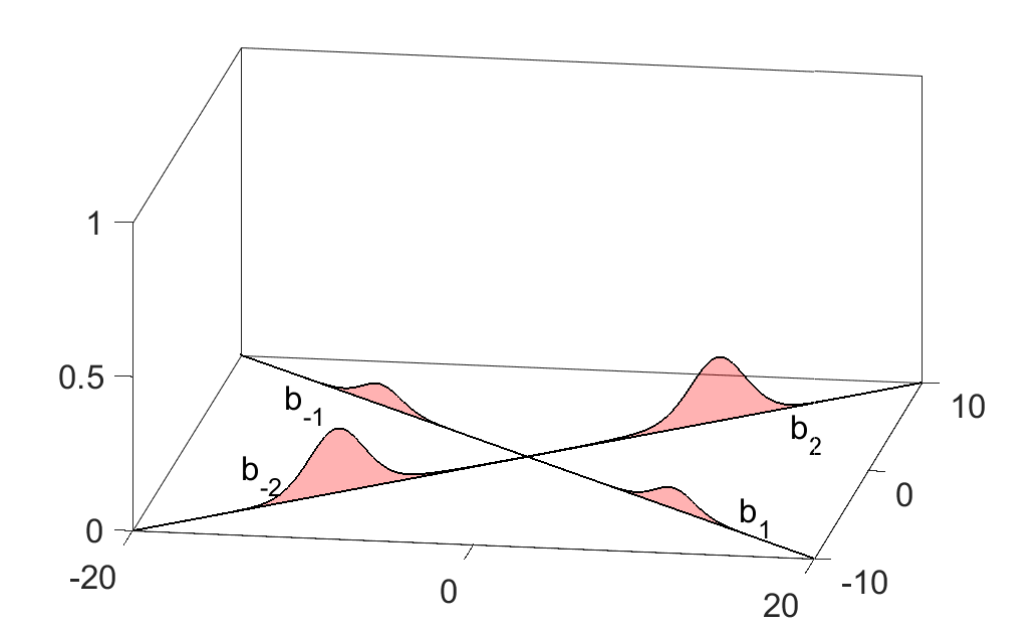}
  \caption{}
  % \label{estcomp}
\end{subfigure}

\caption{Soliton dynamics at different time moments, i.e., $t=0$ (a), $t=0.9$ (b), $t=1.1$ (c) and $t=2$ 
(d) when the sum rule in Eq.~\eqref{constrain1} is fulfilled by choosing the following values of the nonlinearity coefficients: $\beta_{\pm 1}=6$ and $\beta_{\pm 2}=2$.}
\label{fig:sumrule1}
\end{figure}

\begin{figure}[t!]
\includegraphics[width=80mm]{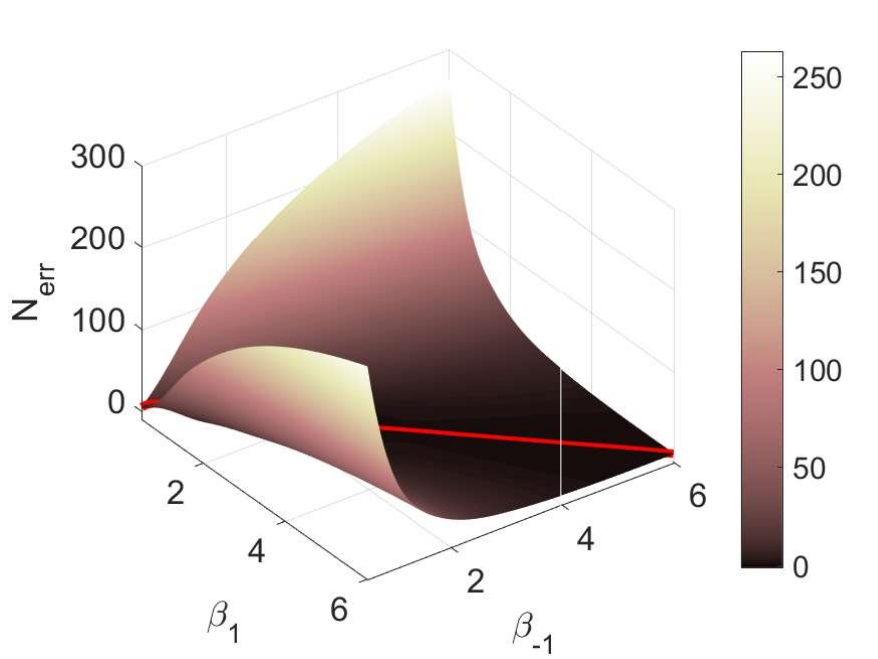}
\includegraphics[width=80mm]{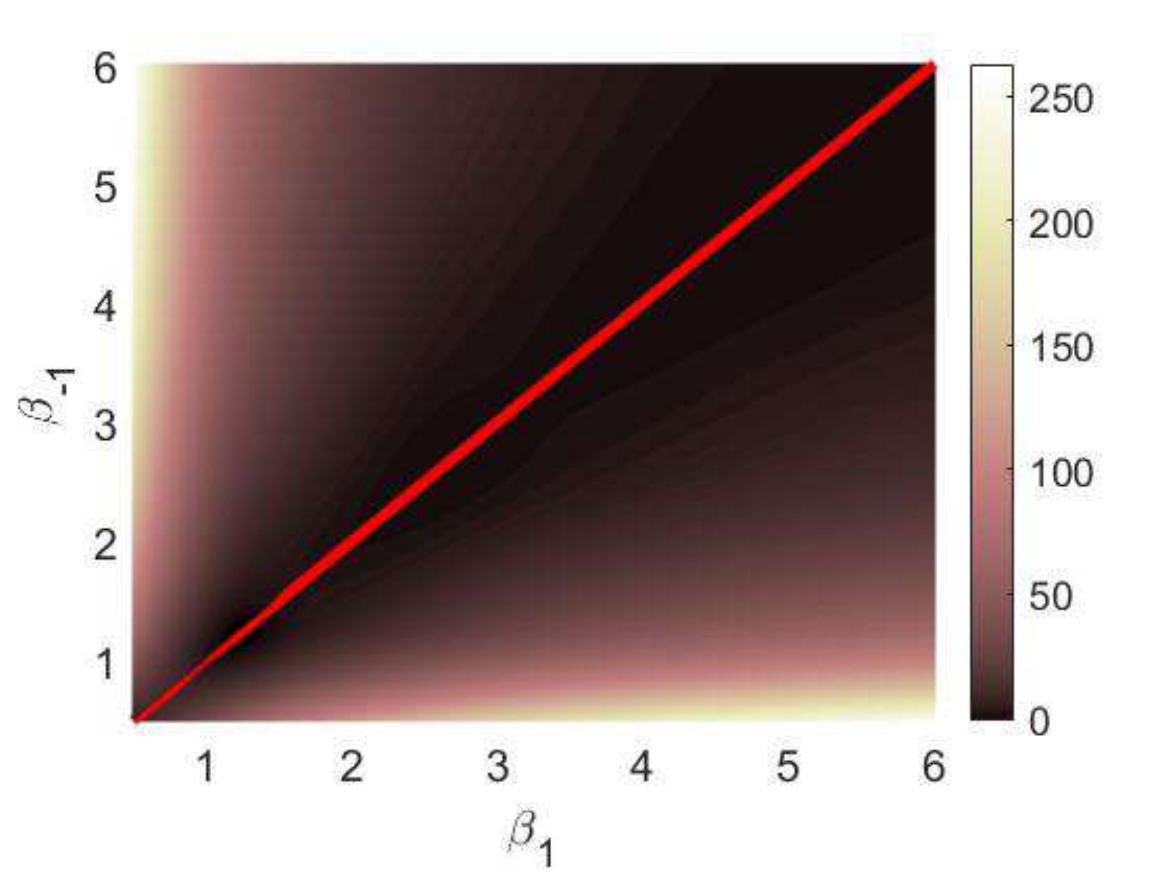}
\caption{Dependence of the norm conservation on values of parameter $\beta_{-1}$ and $\beta_1$ for fixed $\beta_{-2}=\beta_2=2$.} \label{fig:norm}
\end{figure}

% ------- Sum rule 2 is fulfilled in Eq. 38, i.e., tbc is hold
\begin{figure}[hbt!]

\begin{subfigure}{.475\linewidth}
  \includegraphics[width=\linewidth]{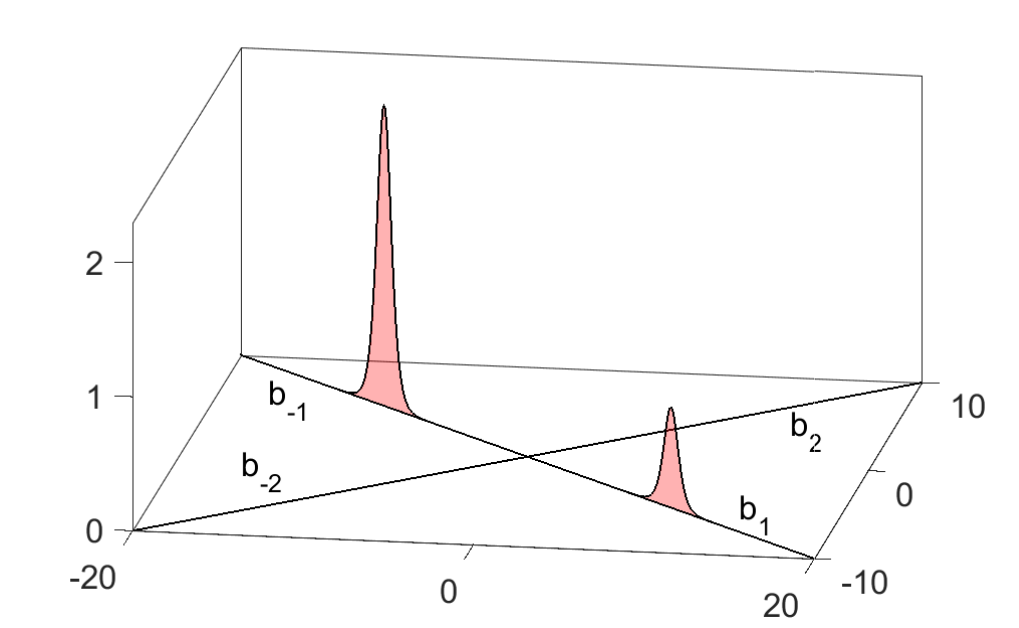}
  \caption{}
  % \label{MLEDdet}
\end{subfigure}\hfill % <-- "\hfill"
\begin{subfigure}{.475\linewidth}
  \includegraphics[width=\linewidth]{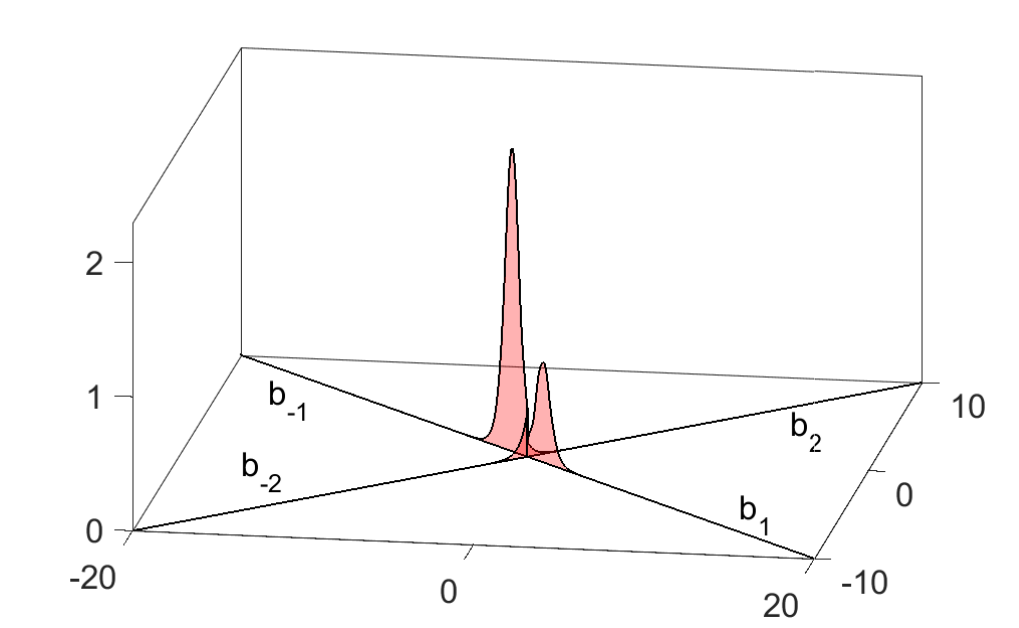}
  \caption{}
  % \label{energydetPSK}
\end{subfigure}

\medskip % create some *vertical* separation between the graphs
\begin{subfigure}{.475\linewidth}
  \includegraphics[width=\linewidth]{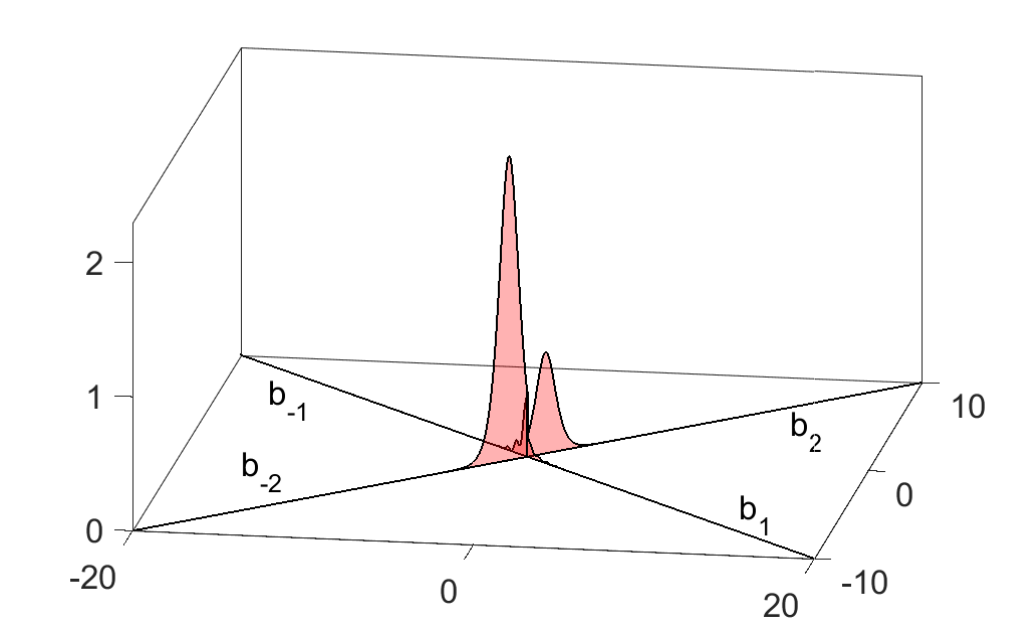}
  \caption{}
  % \label{velcomp}
\end{subfigure}\hfill % <-- "\hfill"
\begin{subfigure}{.475\linewidth}
  \includegraphics[width=\linewidth]{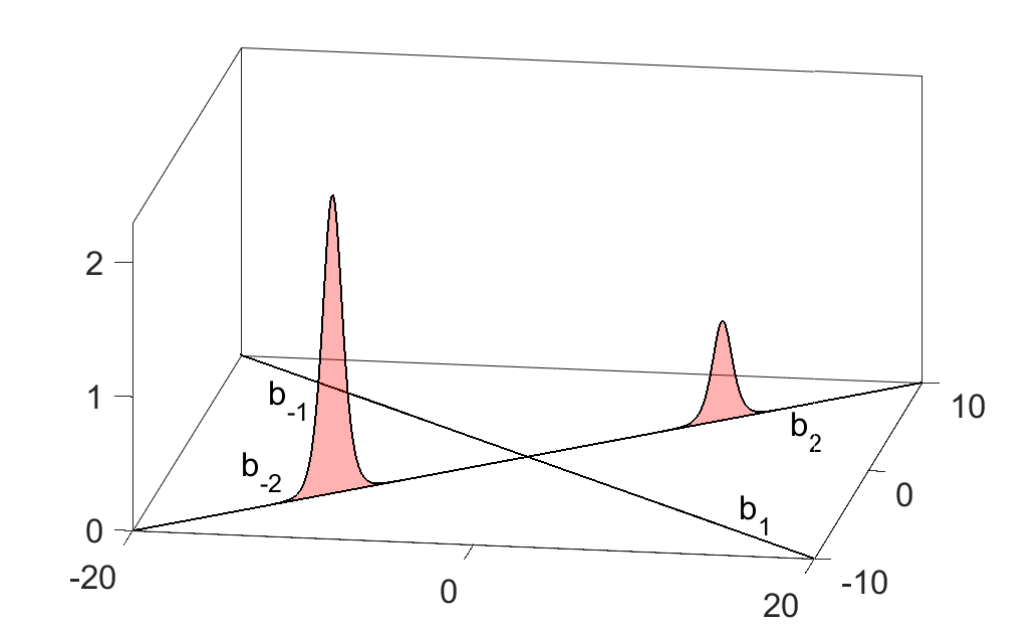}
  \caption{}
  % \label{estcomp}
\end{subfigure}

\caption{Soliton dynamics at different time moments, i.e., $t=0$ (a), $t=0.9$ (b), $t=1.1$ (c) and $t=2$ 
(d) when the sum rule in Eq.~\eqref{sumrule1} is fulfilled by choosing the following values of the nonlinearity coefficients: $\beta_{- 1}=\beta_{- 2}=2$ and $\beta_{1}=\beta_2=6$.}
\label{fig:sumrule2}
\end{figure}

% ------- Sum rules are broken in Eqs. 26 and 38
\begin{figure}[hbt!]

\begin{subfigure}{.475\linewidth}
  \includegraphics[width=\linewidth]{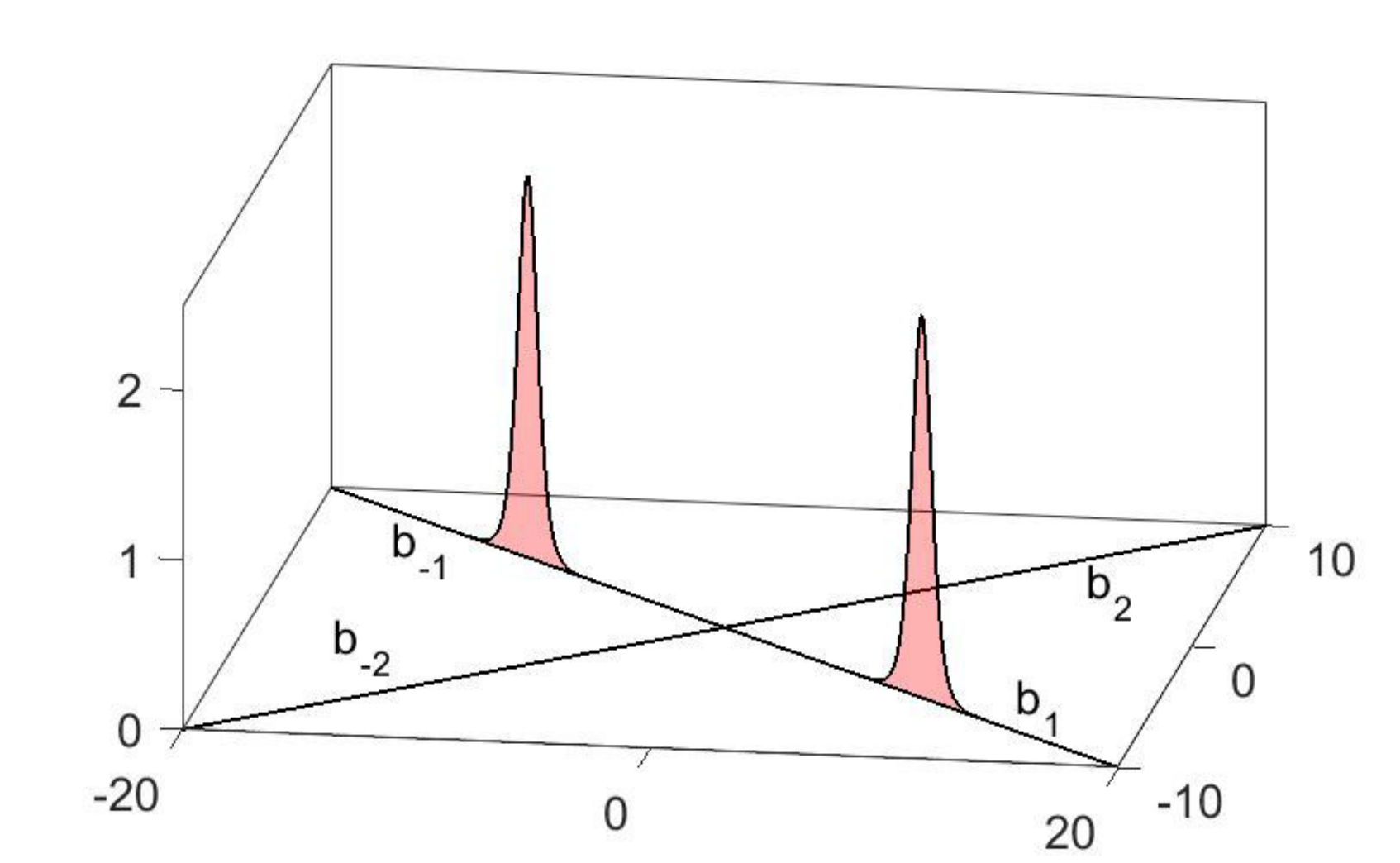}
  \caption{}
  % \label{MLEDdet}
\end{subfigure}\hfill % <-- "\hfill"
\begin{subfigure}{.475\linewidth}
  \includegraphics[width=\linewidth]{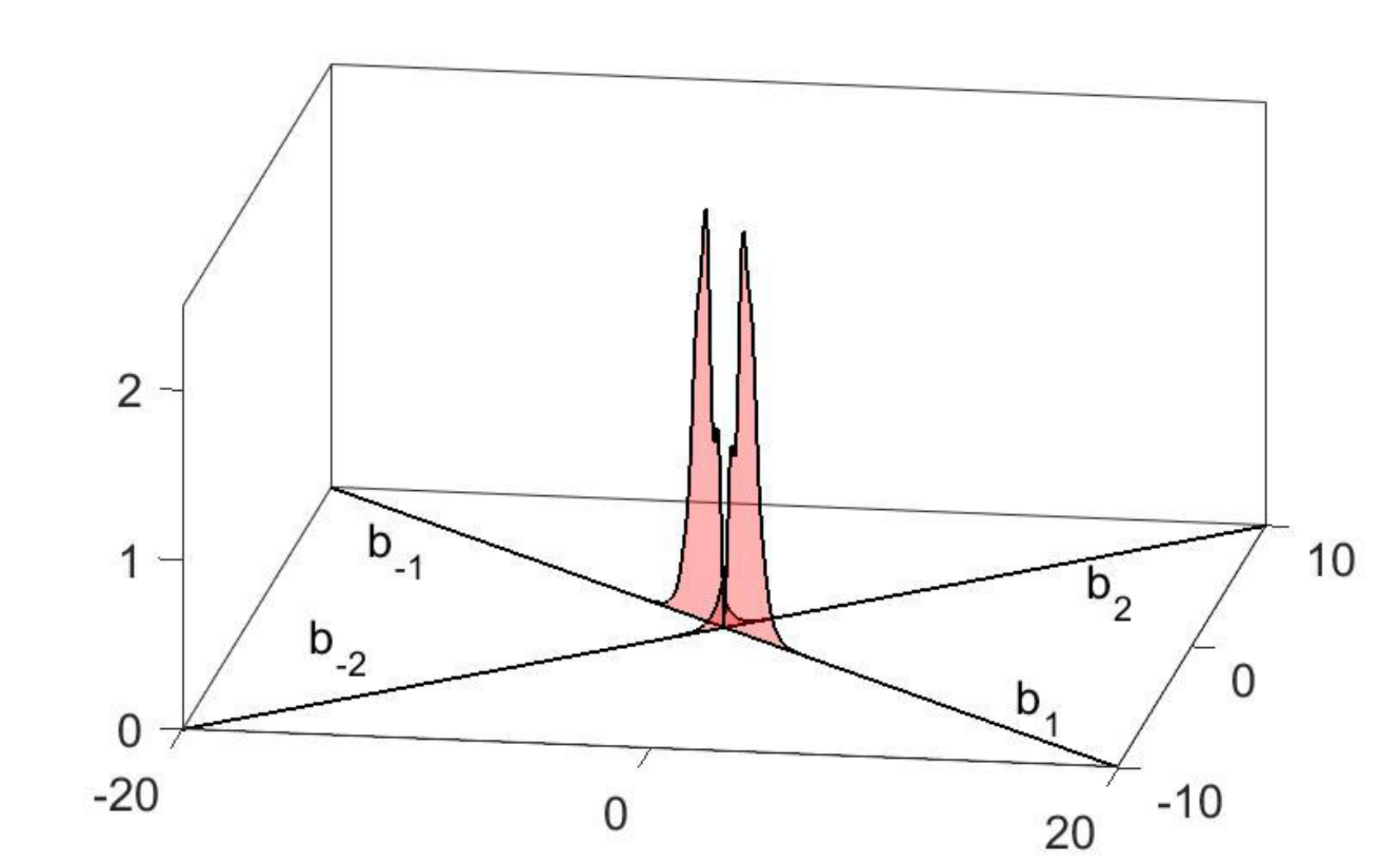}
  \caption{}
  % \label{energydetPSK}
\end{subfigure}

\medskip % create some *vertical* separation between the graphs
\begin{subfigure}{.475\linewidth}
  \includegraphics[width=\linewidth]{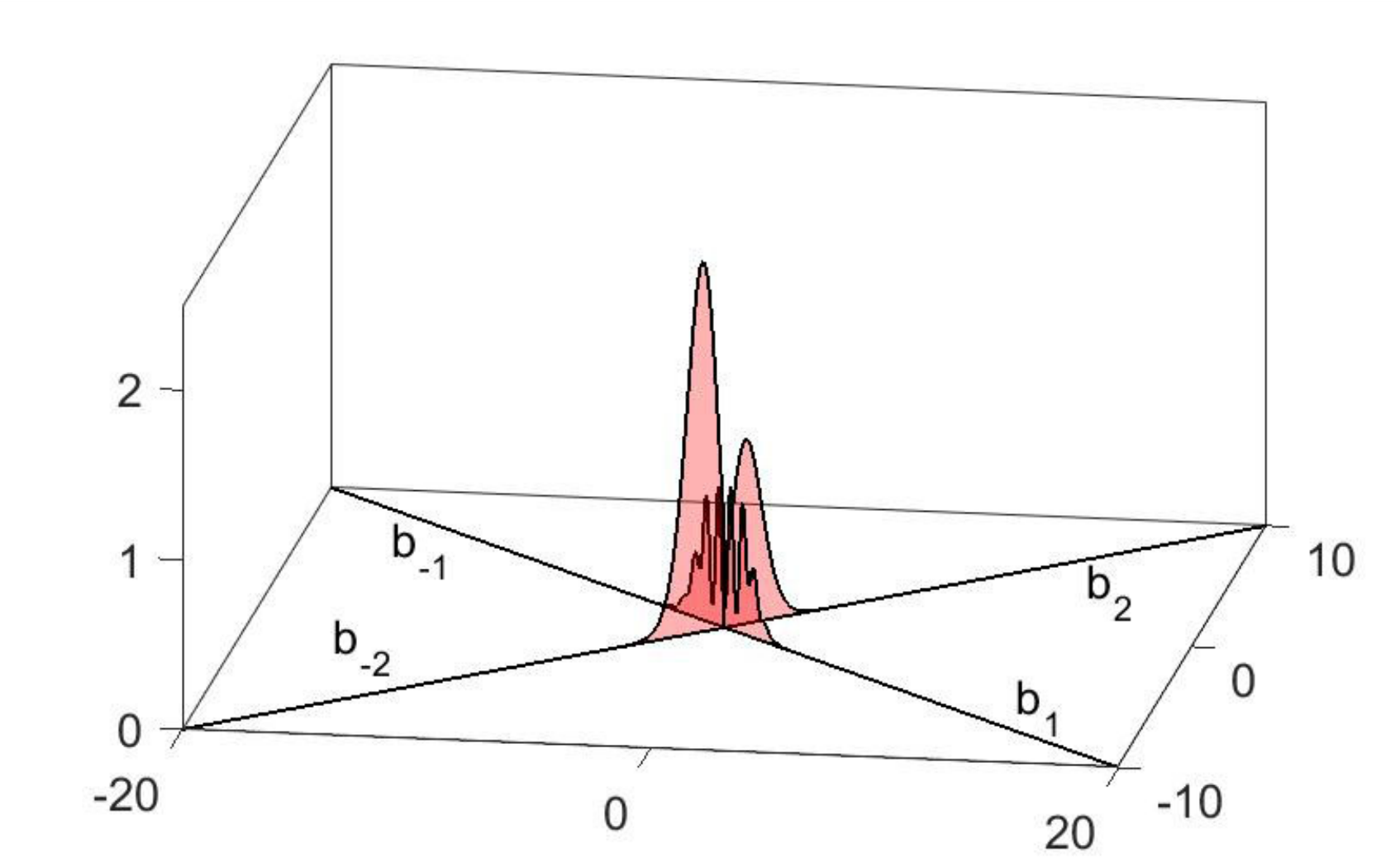}
  \caption{}
  % \label{velcomp}
\end{subfigure}\hfill % <-- "\hfill"
\begin{subfigure}{.475\linewidth}
  \includegraphics[width=\linewidth]{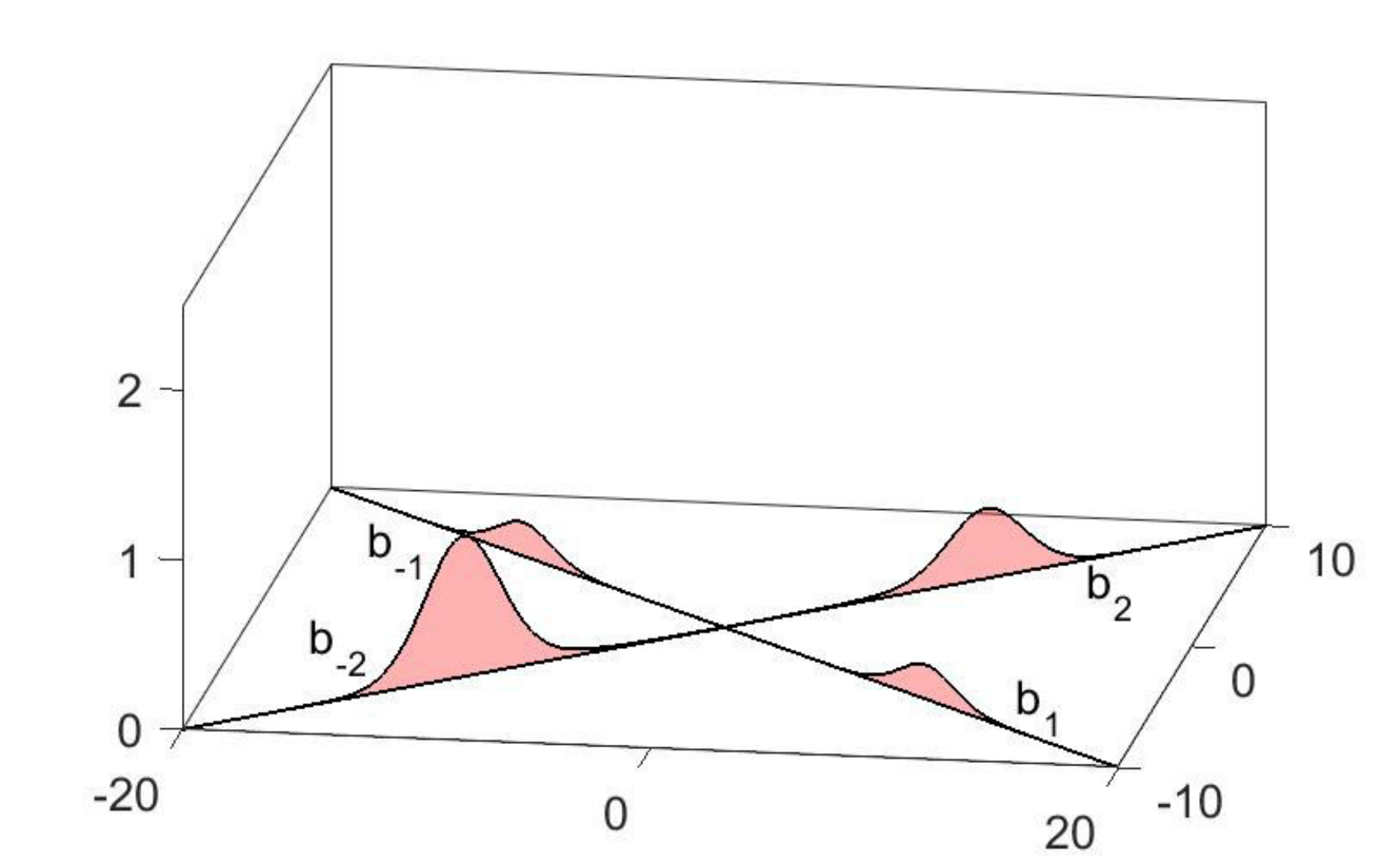}
  \caption{}
  % \label{estcomp}
\end{subfigure}

\caption{Soliton dynamics at different time moments, i.e., $t=0$ (a), $t=0.9$ (b), $t=1.1$ (c) and $t=2$ 
(d) when the sum rules in Eqs.~\eqref{constrain1} and \eqref{sumrule1} are broken by choosing the following values of the nonlinearity coefficients: $\beta_{-1}=\beta_{1}=2$ and $\beta_{-2}=0.5$, $\beta_2=1$.}
\label{fig:broken}
\end{figure}

% \begin{figure}[t!]
% \includegraphics[width=120mm]{contourplot.pdf}
% \caption{Contour plot of soliton dynamics when the sum rule is fulfilled by choosing the following values of the nonlinearity coefficients: $\beta_{\pm j}=2$.} \label{fig:cont_sol_dyn}
% \end{figure}

% \begin{figure}[t!]
% \includegraphics[width=150mm]{broken.pdf}
% \caption{Soliton dynamics at different time moments, i.e., $t=0$ (a), $t=0.8$ (b), $t=1.1$ (c) and $t=2$ 
% (d) when the sum rule is broken by choosing the following values of the nonlinearity coefficients: $\beta_{-1}=2, \; \beta_1=2, \; \beta_{-2}=0.5, \; \beta_2 = 1$.} \label{fig:broken}
% \end{figure}

% \begin{figure}[t!]
% \includegraphics[width=120mm]{contour_broken.pdf}
% \caption{Contour plot of soliton dynamics when the sum rule is broken by choosing the following values of the nonlinearity coefficients: $\beta_{-1}=2, \; \beta_1=2, \; \beta_{-2}=0.5, \; \beta_2 = 1$.} \label{fig:cont_sol_dyn_broken}
% \end{figure}

And the current conservation condition in \eqref{bc1} leads to
\begin{equation}\label{eq3}
    \frac{1}{\sqrt{\beta_{-1}}} (T_0 q_{-1})\big|_{x=0} + 
    \frac{1}{\sqrt{\beta_{1}}} (T_0 q_{1})\big|_{x=0} = 
    \frac{1}{\sqrt{\beta_{-2}}} (T_0 q_{-2})\big|_{x=0} +
    \frac{1}{\sqrt{\beta_{2}}} (T_0 q_{2})\big|_{x=0}.
\end{equation}
% \begin{align}
% \partial_x q_{-1}|_{x=0}&=
% \sqrt{\frac{\beta_{-1}}{\beta_{1}}} \partial_x q_{1}|_{x=0}+
% \sqrt{\frac{\beta_{-1}}{\beta_{2}}} \partial_x q_{2}|_{x=0}-
% \sqrt{\frac{\beta_{-1}}{\beta_{-2}}} \partial_x q_{-2}|_{x=0} \nonumber\\
% &=-\frac{\beta_{-1}}{\beta_{-2}}(T_0 q_{-2})|_{x=0}+
% \frac{\beta_{-1}}{\beta_{1}}(T_0 q_{1})|_{x=0}+
% \frac{\beta_{-1}}{\beta_{2}}(T_0 q_{2})|_{x=0}\nonumber\\
% &=\Bigg[-\frac{\beta_{-1}}{\beta_{1}}+\frac{\beta_{-1}}{\beta_{2}}+\frac{\beta_{-1}}{\beta_{-2}}\Bigg] \left[ e^{-\iu\frac{\pi}{4}}e^{\iu\nu_{-1}}\cdot\partial_t^{1/2}(e^{-\iu\nu_{-1}} q_{-1})|_{x=0}
%      +\iu\frac{1}{4}\partial_x V_{-1}e^{\iu\nu_{-1}}I_t(e^{-\iu\nu_{-1}} q_{-1})|_{x=0}=0 \right].\label{tosumrule}
% \end{align}
% Since the boundary condition in Eq.~\eqref{tbc_qpm1} is required to be transparent, 
% it is true only if the additional factor in Eq.~\eqref{tosumrule} is equal to 1, i.e.\ if the following sum rule is satisfied:
Comparing the above Eqs.~\eqref{eq2} and \eqref{eq3} gives 
\begin{equation}
    \frac{1}{\beta_{-1}}+\frac{1}{\beta_{1}}=\frac{1}{\beta_{-2}}+\frac{1}{\beta_{2}}.
    \label{sumrule1}    
\end{equation}

Thus, fulfilling the sum rule \eqref{sumrule1} implies that the vertex boundary conditions \eqref{bc1} become equivalent to the TBCs at the vertex of the graph. 
In other words, the vertex becomes ``transparent'' with respect to soliton transmission. 
However, since the solution given by Eq.~\eqref{solit} describes a %breathing (non-traveling) 
traveling soliton, such ``transparency'' implies that solitons moving from bonds $b_{\pm 1}$ to bonds $b_{\pm 2}$ transmit through the vertex without any reflection.
Such a property can be demonstrated in the numerical experiments presented in the next section.  
Note that the sum rule \eqref{sumrule1} is different from the one given by \eqref{constrain1} and they coincide only if the parameters $\beta_{\pm j}$ have certain values (for example, if all parameters have the same value, $\beta_{-1}=\beta_{1}=\beta_{-2}=\beta_{2}$, which implies natural boundary conditions). 
This implies that unlike to the case of classical NLS equation \cite{Zarif, Jambul1}, for NNLS equation on metric graphs, integrability is not equivalent to the ``transparency" of the vertex.

%%%%%%%%%%%%%%%%%%%%%%%%%%%%%%%%%%%%%%%%%%%%%%%%%%%%
\section{Numerical experiment}\label{sec:4}
Here we show the results of a numerical experiment performed to verify the results of deriving transparent vertex boundary conditions (TVBCs) for the nonlocal nonlinear Schr\"odinger equation on the star graph shown in Fig.~\ref{fig:star_graph}. 
In this numerical experiment we use Runge-Kutta method. 
In all examples we will use the following initial setup: 
the initial conditions are imposed on $b_{-1}$ and $b_1$ symmetric bonds and chosen as analytical solutions in Eq.~\eqref{travelling_graph}, 
where its parameters are given as $\alpha_1 = 1.13+1.13\iu$, 
$\beta_1 = 1.13-1.13\iu$ (should not be confused with BC parameters)
and $k_1=\pm 2.5+1.5\iu$, $\Bar{k}_1=\mp 2.5+1.5\iu$ for $b_{\pm 1}$ bonds, respectively. 

As a first example we consider the case, when the sum rule \eqref{constrain1} is satisfied. 
The evolution of the traveling solitons for this case is shown in Fig.~\ref{fig:sumrule1} in four consecutive time steps. 
This example can be supported by determining the set of parameters pairs $(\beta_{-1},\beta_1)$ for some fixed $\beta_{-2}=\beta_2$. 
Fig.~\ref{fig:norm} shows the dependence of the deviation of the norm from its mean as a function of the parameters $(\beta_{-1},\beta_1)$.
In this figure you can see the conservation of the norm along the red line.
The deviation of the norm from its mean over the whole time is defined as
\begin{equation}
    N_{\rm err} = \int_0^T |\bar{N}-N|\,dt,
\end{equation}
where 
\begin{equation*}
\bar{N}=\frac{1}{T} \int_0^T N(t)\,dt    
\end{equation*}
is the average value of the total norm over the whole time and $T$ is the total traveling time.

The second example is the case when the sum rule \eqref{sumrule1} is fulfilled, i.e.\ when TBCs are imposed at the central vertex. 
Fig.~\ref{fig:sumrule2} shows the evolution of the traveling solitons for this case in four consecutive time steps. 
The reflectionless transmission of the solitons is evident from this plot.
As a last example, we consider the case where the sum rule is violated. 
The results of the calculations are shown in Fig.~\ref{fig:broken}.   
In this plot one can observe the reflection at the vertex of the graph.

% \TMashrab{The most suitable option can be selected from the Fig. \ref{fig:sumrule2}.}

% \TJambul{In my opinion Fig.\ref{fig:sumrule2} is better. In addition I think we should also include here the same kind of plot for the case when the sum rule is broken. Moreover maybe the case for the sum rule arising from conservative lows should be presented. How do you think?

% Mashrab could you please calculate and put here such a plots? Then we see what to do next.}

% \TMashrab{I have added Figs.~\ref{fig:broken} and \ref{fig:cont_sol_dyn_broken} for the case when the sum rule is not fulfilled.}

\begin{figure}[t!]
\includegraphics[width=80mm]{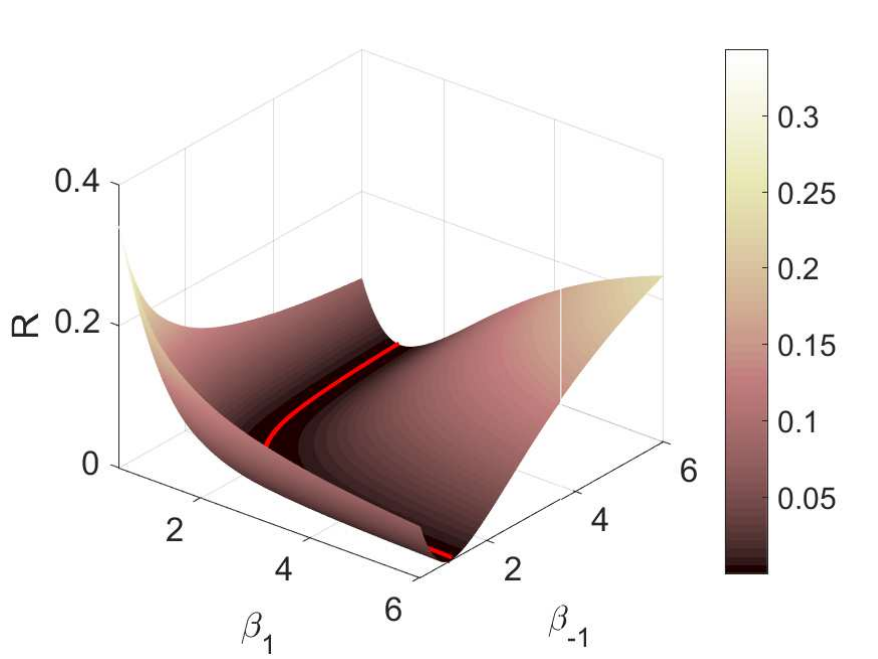}
\includegraphics[width=80mm]{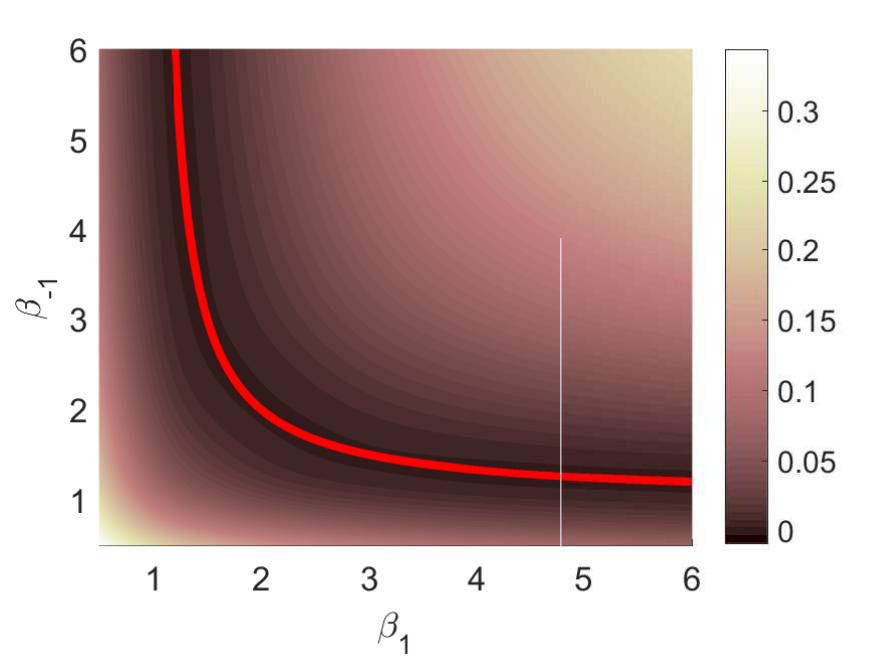}
\caption{Dependence of the reflection coefficient on values of parameter $\beta_{-1}$ and $\beta_1$ for fixed $\beta_{-2}=\beta_2=2$.} \label{fig:refco}
\end{figure}

Note that the choice of parameters $(\beta_{-1},\beta_1)$ is obviously not unique for some fixed $\beta_{-2}$ and ${\beta_2}$. 
This can be verified by plotting the dependence of the reflection coefficient on the parameters $(\beta_{-1}, \beta_1)$.
For some fixed time instant $t_0$, the reflection coefficient can be defined as
\begin{equation}
    R=\frac{N_{-1}+N_1}{N_{-1}+N_1+N_{-2}+N_2},
\end{equation}
where $N_{\pm j}$ are partial norms in Eq.~\eqref{norm1} of bonds $b_{\pm j}$ at time $t_0$. 
The plot of the reflection coefficient as a function of the BC parameters $(\beta_{-1}, \beta_1)$ for fixed $\beta_{-2}=\beta_2=2$ at sufficient time ($t=2$) is shown in Fig.~\ref{fig:refco}.
From this plot one can see the black curve (highlighted by the red line) bounded by the values of the parameters $(\beta_{-1}, \beta_1)$ that satisfy the equation $\beta_{-1}^{-1}+\beta_1^{-1}=1$. 
This shows the manifestation of the reflectionless transition of solitons when the constraint \eqref{sumrule1} is satisfied.

%%%%%%%%%%%%%%%%%%%%%%%%%%%%%%%%%%%%%%%%%%%%%%
% \subsection{Discussion of possible experimental realizations}\label{sec:4}

%%%%%%%%%%%%%%%%%%%%%%%%%%%%%%%%%%%
\section{Conclusions}\label{sec:concl}
In this paper, we have derived transparent boundary conditions for the PT-symmetric nonlocal nonlinear Schr\"odinger equation on metric graphs using the so-called potential approach. 
Constraints that make transparent boundary conditions to weight-continuity and generalized Kirchhoff conditions are derived. 
Numerical utilization of transparent boundary conditions and numerical proof of their equivalence to weight-continuity and generalized Kirchhoff rules are provided. 
For PT-symmetric solitons, the transparency implies the reflectionless transmission of a ``wing" of the soliton from the bond $b_{\pm 1}$ to the bond $b_{\pm 2}$.
The results obtained in this work can be applied to the modeling of optical networks and optoelectronic devices using PT-symmetric solitons, so that the minimum signal loss can be achieved by tuning the soliton propagation.

%%%%%%%%%%%%%%%%%%%%%%%
%\begin{acknowledgments}
\section*{Acknowledgment}
The work is supported by the grant of the Agency for innovative development under the Ministry of higher education, science and innovation of the Republic of Uzbekistan (Ref.\ No.\ F-2021-440) and by the Grant REP-04032022/206, funded under the MUNIS Project, supported by the World Bank and the Government of the Republic of Uzbekistan.
%\end{acknowledgments}

%%%%%%%%%%%%%%%%%%%%%%%%%%%%%%%%%%%% Refs

\end{document}